\documentclass[prl,twocolumn,floatfix,showpacs,superscriptaddress]{revtex4}
\usepackage{graphicx}
\usepackage{amssymb}

\begin{document}
\title{Mesoscopic to universal crossover of transmission phase of
  multi-level quantum dots} 
\author{C.~Karrasch} \affiliation{Institut
  f\"ur Theoretische Physik, Universit\"at G\"ottingen, 37077
  G\"ottingen, Germany} 
\author{T.~Hecht} \affiliation{Physics Department, Arnold
  Sommerfeld Center for Theoretical Physics, and Center for
  NanoScience, \\ Ludwig-Maximilians-Universit\"at, 80333 Munich,
  Germany} 
\author{A.~Weichselbaum} \affiliation{Physics Department, Arnold
  Sommerfeld Center for Theoretical Physics, and Center for
  NanoScience, \\ Ludwig-Maximilians-Universit\"at, 80333 Munich,
  Germany} 
\author{Y.~Oreg} \affiliation{Department of Condensed
  Matter Physics, The Weizmann Institute of Science, Rehovot 76100,
  Israel} 
\author{J.~von Delft} \affiliation{Physics Department,
  Arnold Sommerfeld Center for Theoretical Physics, and Center for
  NanoScience, \\ Ludwig-Maximilians-Universit\"at, 80333 Munich,
  Germany}
\author{V.~Meden} \affiliation{Institut f\"ur
  Theoretische Physik, Universit\"at G\"ottingen, 37077 G\"ottingen,
  Germany} 
\date{March 22, 2007}

\begin{abstract}
  Transmission phase $\alpha$ measurements of many-electron quantum
  dots (small mean level spacing $\delta$) revealed universal phase 
  lapses by $\pi$ between consecutive resonances. In contrast, for 
  dots with only a few electrons (large $\delta$), the appearance 
  or not of a phase lapse depends on the dot parameters. We show 
  that a model of a multi-level quantum dot with local Coulomb 
  interactions and arbitrary level-lead couplings reproduces the 
  generic features of the observed behavior. The universal behavior 
  of $\alpha$ for small $\delta$ follows from Fano-type 
  antiresonances of the renormalized single-particle levels.
\end{abstract}
\maketitle     

One of the longest-standing puzzles in mesoscopic physics is the
intriguing phase-lapse behavior observed in a series of experiments
\cite{ex1,ex2,ex3} on Aharonov-Bohm rings containing a quantum dot in
one arm.  Under suitable conditions in linear response, both the phase
and magnitude of the transmission amplitude $T = |T|e^{i \alpha}$ of
the dot can be extracted from the Aharonov-Bohm oscillations of the
current through the ring.  If this is done as function of a plunger
gate voltage $V_g$ that linearly shifts the dot's single-particle
energy levels downward, $\varepsilon_j = \varepsilon_j^0 - V_g$
($j=1,2, \dots$ is a level index), a series of well-separated
transmission resonances [peaks in $|T(V_g)|$, to be called ``Coulomb
blockade'' (CB) peaks] of rather similar width and height was
observed, across which $\alpha (V_g)$ continuously increased
by $\pi$, as expected for Breit-Wigner-like resonances.  In each CB
valley between any two successive CB peaks, $\alpha$ always jumped
sharply downward by $\pi$ (``phase lapse'', PL).  The PL behavior was
observed to be ``universal'', occurring in a large succession of
valleys for every many-electron dot studied in~\cite{ex1,ex2,ex3}.
This universality is puzzling, since naively the behavior of $\alpha
(V_g)$ is expected to be ``mesoscopic'', \emph{i.e.} to show a PL in
some CB valleys and none in others, depending on the dot's shape, the
parity of its orbital wavefunctions, etc.  Despite a large amount of
theoretical work (reviewed in \cite{Hackenbroich,Gefen02}), no fully 
satisfactory framework for
understanding the universality of the PL behavior has been found yet.

A hint at the resolution of this puzzle is provided by the most recent
experiment \cite{ex3}, which also probed the few-electron regime: as
$V_g$ was increased to successively fill up the dot with electrons,
starting from electron number $N_{\rm e} = 0$, $\alpha (V_g)$ was
observed to behave mesoscopically in the \emph{few}-electron regime,
whereas the above-mentioned universal PL behavior emerged only in the
\emph{many}-electron regime $(N_{\rm e} \gtrsim 15)$.  Now, one
generic difference between few- and many-electron dots is that the
latter have smaller level spacings $\delta_j = \varepsilon_{j+1}^0 -
\varepsilon_j^0$ for the topmost filled levels.  With increasing
$N_{\rm e}$, their $\delta_j$'s should eventually become smaller than
the respective level widths $\Gamma_j$ stemming from hybridization
with the leads. Thus, Ref.~\onlinecite{ex3} suggested that a key
element for understanding the universal PL behavior might be that
\emph{several overlapping single-particle levels simultaneously}
contribute to transport. Due to the dot's Coulomb charging energy
$U$, the transmission peaks remain well separated nevertheless.

Previous works have studied 
the transmission amplitude of multi-level, interacting dots 
\cite{Bruder,OregGefen97,Imry,VF,Gefen}. 
However, no systematic study has yet been performed 
of the interplay 
of level spacing, level widths, and charging energy that combines 
a wide range of parameter choices with an accurate treatment of 
the correlation effects induced by the Coulomb interaction. 
The present Letter aims to fill this gap by using two powerful 
methods, the numerical (NRG)
\cite{Krishna-murthy,Weichsel} and functional (fRG) \cite{TCV} 
renormalization group approaches, to study
systems with up to 4 levels (for spinless electrons; see below).
We find that if the ratio
of average level spacing $\delta$
to average level width $\Gamma$ is decreased into the regime
$\delta \lesssim \Gamma$, one of the \emph{renormalized effective 
single-particle levels generically} becomes wider than all 
others, and hovers in the vicinity of the chemical potential $\mu$ 
in the regime of $V_g$ for which the PLs occur. 
Upon varying $V_g$, the narrow 
levels cross $\mu$ \emph{and} the broad level, leading 
to Fano-type antiresonances accompanied by universal PLs. 
For $\delta \gtrsim \Gamma$, $\alpha(V_g)$ behaves mesoscopically
\cite{Silva} for all $U$.
Decreasing $\delta$ thus causes the PL behavior 
to generically change from mesoscopic to universal, 
as observed experimentally \cite{ex3}. 

\emph{Model:} The dot part of our model Hamiltonian is 
\vspace{-2.7mm}
\begin{eqnarray*}
\label{dotham}
H_{\rm dot} = \sum_{j=1}^N \varepsilon_j n_{j} + {1 \over 2} U
\sum_{j \neq j'} \left(n_{j}-{\textstyle \frac{1}{2}} \right)
\left(n_{j'}-{\textstyle \frac{1}{2}}\right) \, , 
\end{eqnarray*}
with $n_{j} = d_{j}^\dag d_{j}$ and dot creation operators
$d_{j}^\dag$ for spinless electrons, where $U>0$ describes Coulomb
repulsion. The semi-infinite leads are modeled by a tight-binding
chain $H_l = -t \sum_{m=0}^{\infty} (c^\dag_{m,l}c_{m+1,l} +
\mbox{H.c.})$ and the level-lead couplings by $H_T=- \sum_{j,l} (t_j^l
c^\dag_{0,l} d_{j} + \mbox{H.c.})$, where $c_{m,l}$ annihilates 
an electron on site $m$ of lead $l=L,R$ and $t_j^l$ are real
level-lead hopping matrix elements.  Their relative signs for
successive levels, $s_j=\mbox{sgn}(t_j^L t_j^R t_{j+1}^L t_{j+1}^R)$,
are sample-dependent random variables determined by the parity of the
dot's orbital wave functions. The effective width of level $j$ is
given by $\Gamma_j = \Gamma_j^L+\Gamma_j^R$, with $\Gamma_j^l = \pi
\rho |t_j^l|^2 $. We take $\rho$, the local density of states at
the end of the leads, to be energy independent, choose $\mu = 0$, 
and specify our choices of $t_j^l$ using the
notation $\sigma = \{s_1, s_2, \dots \}$,
$\gamma=\{\Gamma_1^L,\Gamma_1^R,\Gamma_2^L,\ldots\}$, $\Gamma =
{1\over N} \sum_{j,l} \Gamma_j^l$.

\emph{Methods:} We focus on linear response transport and 
unless stated otherwise, on zero temperature $(\tau = 0)$. 
Then the dot produces purely elastic, potential
scattering between left and right lead, characterized by the
transmission matrix $T_{ll'} = 2 \pi \rho \sum_{ij}t_i^l {\cal
  G}^R_{ij} (0) t_j^{l'}$, where ${\cal G}^R_{ij}(\omega)$ is the
retarded local Green function which we compute using NRG and fRG. 
The NRG is a numerically exact method that is known to produce 
very accurate 
results \cite{Krishna-murthy,Weichsel}. 
The fRG is a renormalization procedure for the self-energy $\Sigma$ 
and higher order vertex functions (see~\cite{TCV} for details).  
We use a truncation scheme that keeps the flow equations for $\Sigma$ and
for the frequency independent part of the effective two-particle
(Coulomb) interaction.  Comparisons with NRG \cite{TCV} have shown
this approximation to be reliable provided that the number of 
(almost) degenerate levels and the interaction do not become too 
large. fRG is much cheaper computationally than NRG, enabling 
us to efficiently explore  the vast parameter space relevant for 
multi-level dot models.

At the end of the fRG flow, the full Green function at zero frequency
takes the form $[{\cal G}^R(0)]^{-1}_{ij} = - h_{ij} + i \Delta_{ij}$,
with an effective, \emph{noninteracting} (but $V_g$ \emph{and} $U$-dependent)
single-particle Hamiltonian $h_{ij} = (\varepsilon_j^0 - V_g)
\delta_{ij} + \Sigma_{ij}$, whose level widths are governed by
$\Delta_{ij} = \pi \rho \sum_l t_i^l t_j^l$.
To interpret our results, we  
adopt the eigenbasis of $[{\cal G}^R(0)]^{-1}_{ij}$,
with eigenvalues $-\tilde \varepsilon_j + i \tilde \Gamma_j$, 
and view $\tilde \varepsilon_j$ and $\tilde \Gamma_j$ as 
level positions and widths of a renormalized effective 
model (REM) describing the system.

\begin{figure}[t]
\begin{center}
\includegraphics[width=.45\textwidth,clip]{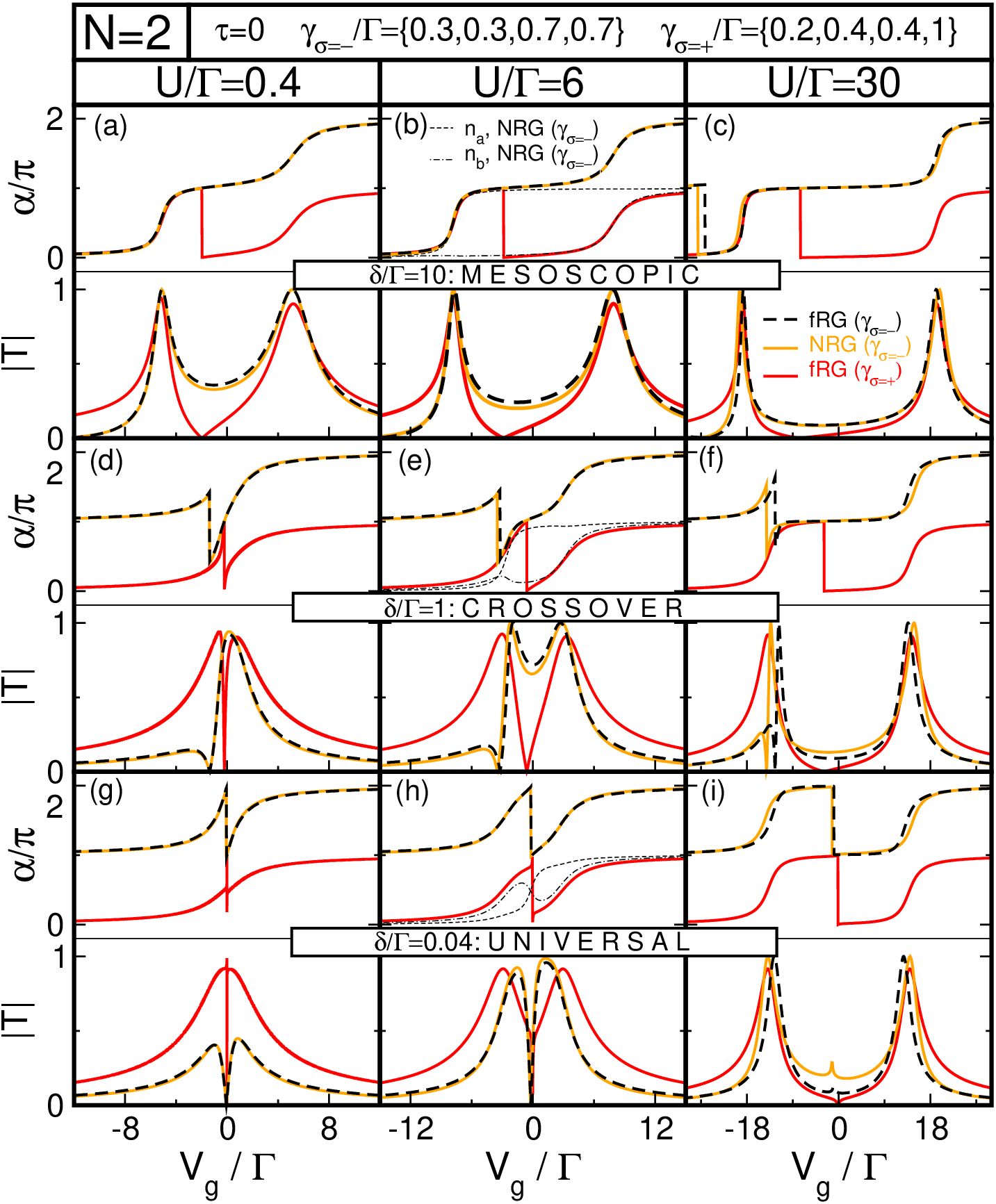} 
\end{center}
\vspace{-4ex}
\caption[]{$|T(V_g)|$ and $\alpha (V_g)$ for $N=2$,
  $\varepsilon_{2,1}^0 = \pm \delta/2$, and $\tau=0$: decreasing $\delta/\Gamma$
  produces a change from (a,b,c) mesoscopic via (d,e,f) crossover to
  (g,h,i) universal behavior; increasing $U/\Gamma$ leads to increased
  transmission peak spacing. (b,e,h) include the occupancies $n_+$
  (thin dashed) and $n_-$ (thin dash-dotted) of the levels coupled to
  the even/odd lead in the case of LR-symmetry. The condition 
 $n_+ =  n_- \, \mbox{mod} \, 1$ 
produces a TZ and PL.  
For the blip and hidden TZ near $V_g=0$ in (i), 
see \cite{footnoteCIR}.}
\label{fig1}
\vspace{-5ex}
\end{figure}

For LR-symmetry, $\Gamma^L_j = \Gamma^R_j$, an NRG-shortcut can be
used, which is much less demanding than computing the full 
$G^R_{ij}(\omega)$: the S-matrix is then
diagonal in the even-odd basis of the leads and its 
eigenvalues depend on the total occupancies $n_\pm$ of all levels coupled
to the even/odd lead (Friedel sum rule), 
so that the transmission amplitude $T = T_{LR}$ takes the 
form $T = \sin [\pi (n_+ - n_-)] e^{i \pi (n_+ + n_-)}$.
A transmission zero (TZ) and hence PL occurs when 
$n_+ =  n_- \, \mbox{mod} \, 1$
[Figs.~\ref{fig1}(e,h): $n_\pm$ in thin dashed/dashed-dotted line].

\begin{figure*}[t]
\vspace{-3ex}
\begin{center}
\includegraphics[width=.89\textwidth,clip]{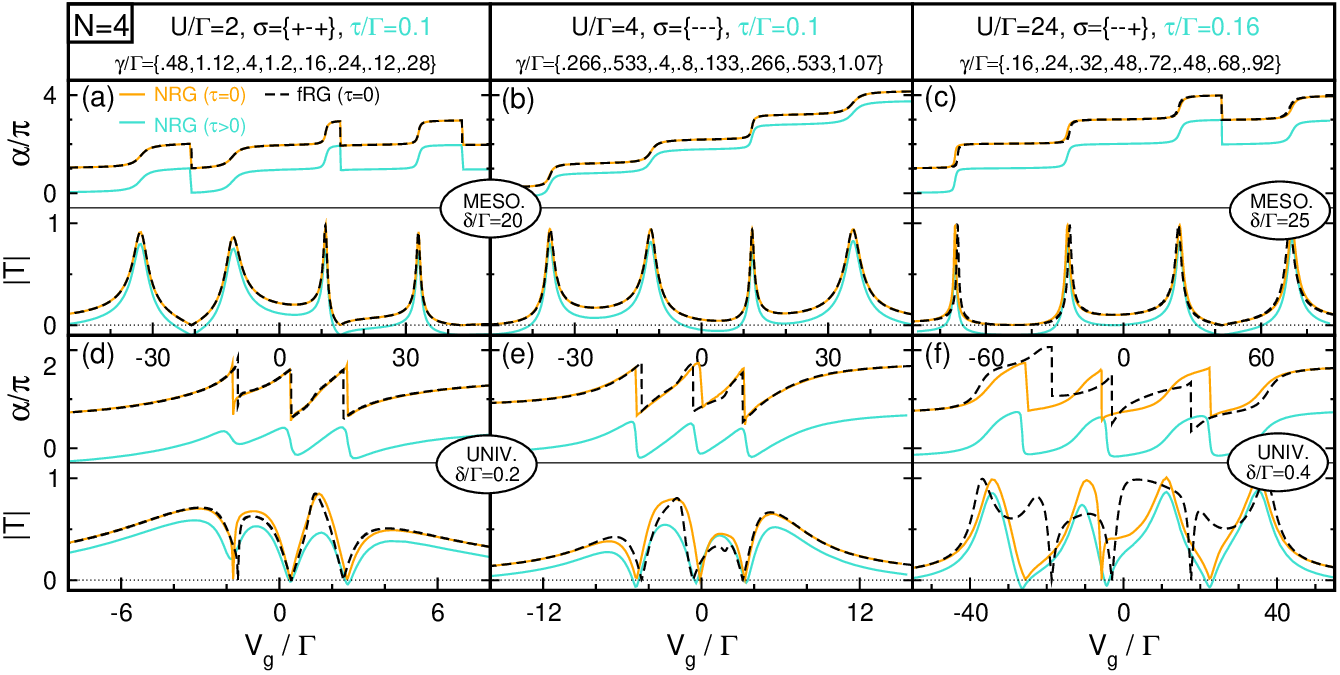} 
\end{center}
\vspace{-5ex}
\caption[]{$|T(V_g)|$ and $\alpha (V_g)$ for $N=4$, with equidistant levels, 
  $\delta_j \equiv \delta$.
  The qualitative features do not change if this assumption is
  relaxed, or if $U$ is assumed to be slightly level-dependent, $U \to
  U_{jj'}$.  Decreasing $\delta/\Gamma$ produces a crossover from
  (a,b,c) mesoscopic to (d,e,f) universal behavior; increasing
  $U/\Gamma$ leads to increased spacing of the transmission peaks and PLs.
  For clarity the finite temperature ($\tau>0$) curves were shifted 
  downwards (by $0.1$ for $|T|$). 
  (f) For $U/\Gamma\gg 1$ and $\tau/\Gamma=0.16$, the CB peak and PL 
   shapes are strikingly 
   similar to those observed experimentally at comparable ratios of 
   $\tau/\Gamma$  (see Fig.~3 of \cite{ex2} and Fig.~6 
  of \cite{ex3}).  For small $\delta/\Gamma$ \emph{and} 
  $U/\Gamma \gg 1$ fRG becomes less reliable and the results begin to
  differ from those of NRG.
\label{fig2}}
\vspace{-4ex}
\end{figure*}

\begin{figure}[t]
\vspace{-3ex}
\begin{center}
\includegraphics[width=.424\textwidth,clip]{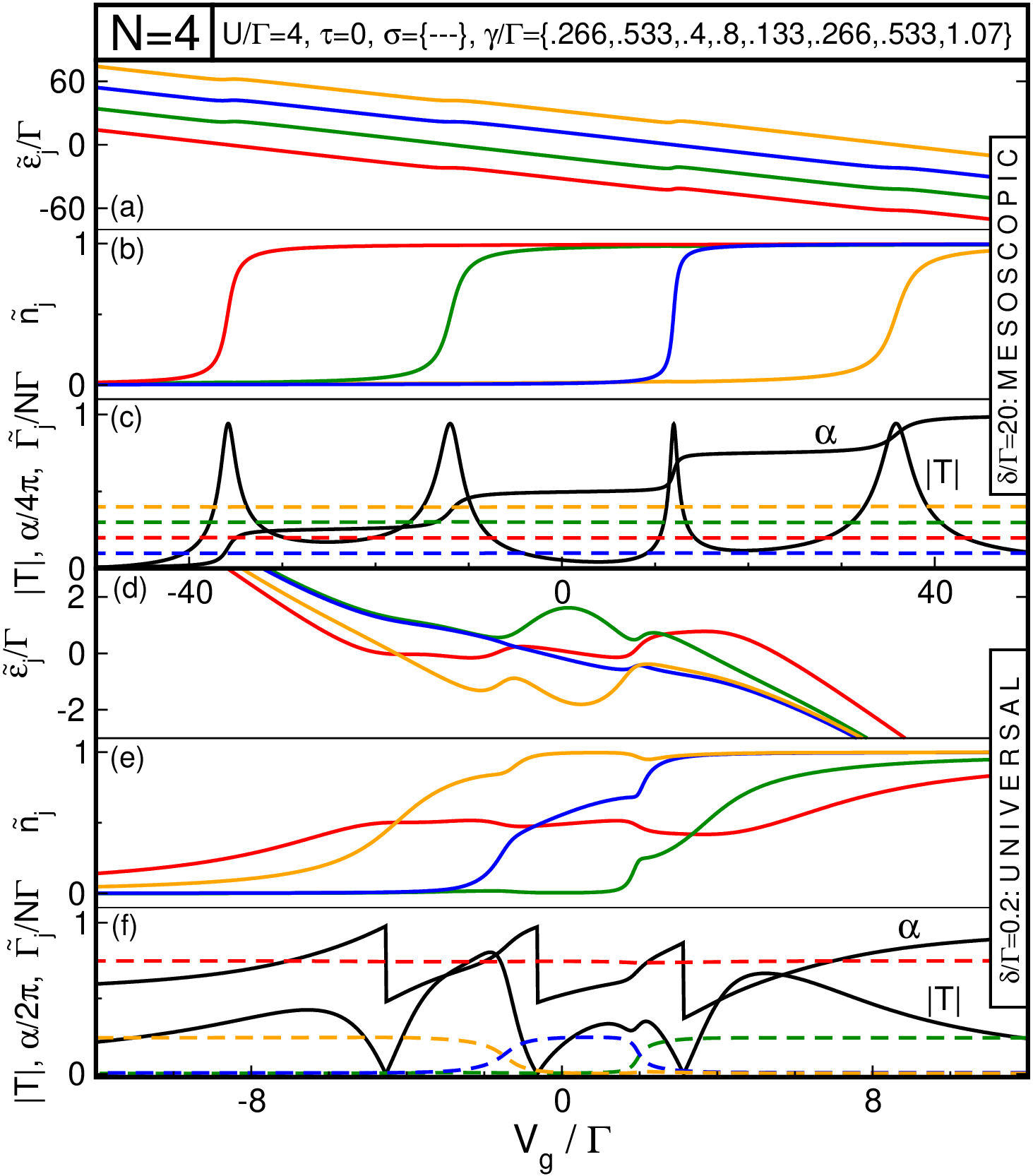} 
\end{center}
\vspace{-4ex}
\caption[]{Renormalized single-particle energies 
$\tilde \varepsilon_j$, occupancies $\tilde n_j$ and 
level widths $\tilde \Gamma_j$ (dashed) of the REM, 
and the resulting $|T|$ and $\alpha$ (all as functions of $V_g$), for 
$N=4$ and $\delta_j \equiv \delta$ at $\tau=0$. 
The parameters are the same as in Fig.~\ref{fig2}(b,e).\label{fig3}}
\vspace{-4ex}
\end{figure}

\emph{Results:} Our results are illustrated in Figs.~\ref{fig1}
to \ref{fig3}. 
fRG and NRG data generally coincide rather well (compare black 
and orange lines in Figs.~\ref{fig1} and \ref{fig2}), 
except for $N=4$ when both $U \gg \Gamma $, $\delta < \Gamma$, 
and correlations become very strong [Fig.~\ref{fig2}(f)]. 
The figures show the following striking qualitative features,
that we found to be generic by running the fRG for ten-thousands of 
parameter sets, which is possible as a complete $T(V_g)$ curve 
can be obtained within a few minutes on a standard PC: 

\emph{Mesoscopic regime:} For $\delta \gtrsim \Gamma$ 
[Figs.~\ref{fig1}(a,b,c), \ref{fig2}(a,b,c)], 
we recover behavior that is similar to the $U=0$ case.
Within the REM it can be understood as transport occurring through
only one effective level at a time [see Fig.~\ref{fig3}(a,b,c)],
with $\tilde \Gamma_j \simeq \Gamma_j$. Each $\tilde \varepsilon_j$ 
that crosses $\mu$ produces a Breit-Wigner-like transmission resonance of 
width $2 \Gamma_j$ and height governed by $\Gamma_j^{L}/\Gamma_j^R$.
At the crossing the other levels are shifted upwards by $U$ [charging
effect; Fig.~\ref{fig3}(a)] leading to renormalized peak 
separations (``level spacings'') $\delta_j+U$ . Between two peaks, 
$\alpha (V_g)$ behaves mesoscopically: depending 
on the sign $s_j$ one either observes a PL
($s_j=+$) or continuous evolution of $\alpha$ ($s_j=-$)
\cite{Silva}.  Additional PLs occur to the left or right, beyond the
last transmission peak [Fig.~\ref{fig2}(a)].  

\emph{Mesoscopic to universal crossover:} As the ratio $\delta /
\Gamma$ is reduced, the behavior changes dramatically: the TZs and PLs
that used to be on the far outside move inward across CB resonances
[see evolution in Figs.~\ref{fig1}(b,e,h)].

\emph{Universal regime:} A universal feature \cite{N=2special} emerges 
for $\delta \lesssim \Gamma \lesssim U$ 
(crossover scales are of order 1, but depend  on
the chosen parameters): 
for \emph{all} choices of the signs $\sigma$ and generic
couplings $\gamma$, the $N$ CB peaks over which $\alpha$
increases by $\pi$ \emph{are separated by $N-1$ PLs,} each accompanied
by a TZ [Figs.~\ref{fig1}(g,h,i), \ref{fig2}(d,e,f)].  This is
consistent with the experimentally observed trend.  For
small to intermediate values of $U/\Gamma$ [Figs.~\ref{fig1}(g,h),
\ref{fig2}(d,e)], the transmission peaks are not well-separated, and
$\alpha (V_g)$ has a saw-tooth shape.  
As $U/\Gamma$ increases so does the peak separation 
and the corresponding phase rises take a more
$S$-like form [Figs.~\ref{fig1}(h,i), \ref{fig2}(e,f)]. 
At finite temperatures of order $\tau \gtrsim \delta$ \cite{footnotetau}
sharp features are smeared out (Fig.\ \ref{fig2}). For 
$U/\Gamma$ as large as in Figs.~\ref{fig1}(i) and \ref{fig2}(f), 
the behavior of $\alpha (V_g)$ (both the $S$-like rises and the universal
occurrence of PLs in each valley) as well as the one of 
$|T(V_g)|$ (similar width 
and height of all CB peaks) is very reminiscent of that observed
experimentally, in particular for $\tau \gtrsim \delta$ [Fig.\
\ref{fig2}(f)].  
For $U/\Gamma \gg 1$ the full width of 
the CB peaks is of order $2N\Gamma$ (not $2\Gamma_j$ as in the
mesoscopic regime), indicating that several 
bare single-particle levels simultaneously contribute to 
transport. The $\tau$ dependence of the width of the PLs
is different from the behavior $\tau^2/(\delta+U)^2$ found in the
mesoscopic regime \cite{twolevelpaper} and will be discussed
in an upcomming publication. Note that for the 
temperatures considered here the width of the PLs is still 
much smaller than the width of the CB peaks.   

For certain fine-tuned parameters ($\gamma$ and $\sigma$) 
the behavior at small $\delta/\Gamma$ deviates from the generic 
case.  For $N=2$ the nongeneric cases were 
classified in~\cite{VF}. In Fig.~\ref{fig1} only generic 
parameters are shown. For $N \geq 3$ LR-symmetric 
couplings produce nongeneric features. 
However, these features are irrelevant to the experiments. They
quickly disappear upon switching on LR-asymmetry or $\tau >0$. 

\emph{Interpretation:} We can gain deeper insight into the appearance
of the TZs and PLs in the universal regime from the properties of the
REM obtained by fRG for moderate $U/\Gamma$ [at which NRG and fRG
agree well; Fig.~\ref{fig2}(b,e)]. 
For $N\geq 3$, $\delta \lesssim \Gamma$ and $U=0$, two of the effective
levels are much wider than the others,
since $\Delta_{ij}$, being a matrix of rank 2, has only two 
nonzero eigenvalues \cite{Thouless} (for the $N=2$ case, see
\cite{N=2special,twolevelpaper}). 
We found that this also holds at $U>0$: for $N =3,4$ one effective 
level is typically a factor of 2 to 3 
wider than the second widest, while the remaining 1 or 2 levels are 
very narrow [Fig.~\ref{fig3}(f)]. At 
$\delta \lesssim \Gamma$ [Fig.~\ref{fig3}(d)] the \emph{interaction} leads
to a highly nonmonotonic dependence of $\tilde \varepsilon_j$ on
$V_g$ 
which is essential for our universal PL scenario: As $V_g$ is swept,  
the widest level hovers in the vicinity of $\mu$ over an extended 
range of $V_g$ values, whereas the narrow ones cross $\mu$ 
-- and therefore also the widest one -- rather rapidly.  This leads to 
a Fano-type effect \cite{Fano,Clerk,Entin-WohlmanLevinson02,Aikawa,Yuval} 
whose effective Fano parameter $q$ is real, by time-reversal
symmetry \cite{Clerk}. Thus, each TZ, and hence PL, can be 
understood as a Fano-type antiresonance arising (irrespective of
the signs of $t_i^l$) from destructive interference
between transmission through a wide and a narrow level.  
The crossings of the narrow levels and $\mu$, and thus the PLs, 
are separated by $ U$ due to charging effects. 
In contrast, for $U=0$, $\tilde \varepsilon_j \propto -V_g$ for 
\emph{all} renormalized levels and no levels cross each other.
Our fRG studies indicate that for the regime $\delta \lesssim
  \Gamma$, the Fano-antiresonance mechanism is generic for 
$U \gtrsim \Gamma$.
We thus expect it 
to apply also for interactions $U \gg \Gamma $ for which fRG is no
longer reliable.
 
The fact that the combination of a wide and several narrow
levels leads to PLs was 
first emphasized in \cite{Imry} (without reference to Fano physics). 
However, whereas in \cite{Imry} a \emph{bare}
wide level was introduced 
as a model assumption (backed by numerical simulations for noninteracting dots
of order 100 levels), in our case a \emph{renormalized} wide level is generated
for \emph{generic} couplings if $\delta \lesssim \Gamma$. 
Also, whereas in \cite{Imry} the wide level repeatedly
empties into narrow ones as $V_g$ is swept (because 
$\Gamma_{\rm wide} \ll U$  was assumed), this strong
occupation inversion \cite{Koenig} is not required in 
our scenario. In Fig.~\ref{fig3}(e), e.g., the wide level remains roughly
half-occupied for a large range of $V_g$, but TZs and PLs occur nevertheless.
We thus view occupation inversion, if it occurs, as a side effect, 
instead of being the cause of PLs \cite{U0remark}. 

\emph{Conclusions:} The most striking feature of our results, based on
exhaustive scans through parameter space for $N=2,3,4$, is that for
any given, generic choice of couplings ($\gamma$ and $\sigma$), the
experimentally observed crossover \cite{ex3} from mesoscopic to
universal $\alpha (V_g)$-behavior can be achieved within our model by
simply changing the ratio $\delta/\Gamma$ from $\gtrsim 1$ to
$\lesssim 1$, provided that $U \gtrsim \Gamma$.  
The universal $\pi$ PLs result from Fano-type
antiresonances of effective, renormalized levels, which arise 
because interactions cause a broad level 
(occuring, if $\delta \lesssim \Gamma$, already for $U=0$)
to be repeatedly crossed by narrow levels. 
A quantitative description requires correlations to be treated 
accurately. 
We expect that the main
features of this mechanism carry over to the case of
spinful electrons, since for 
$\delta \lesssim \Gamma$ spin 
correlation physics (such as the Kondo effect) does not play 
a prominent role \cite{twolevelpaper2}. 

We thank P.\ Brouwer, Y.\ Gefen, L. Glazman, D.\ Golosov, M.\ Heiblum,
J.\ Imry, J.\ K\"onig, F.\ Marquardt, M.\ Pustilnik, H.\ Schoeller,
K.\ Sch\"onhammer, and A.\ Silva for valuable discussions.  This work
was supported in part by the DFG, for VM by SFB602; for TH and JvD by
SFB631, Spintronics RTN (HPRN-CT-2002-00302), NSF (PHY99-07949) and
DIP-H.2.1; and for YO by DIP-H.2.1, BSF and the Humboldt foundation.


\begin{thebibliography}{*}
\bibitem{footnoteCIR} For $N$\,=\,2, $U$ larger than a critical value
  and in the limit $\delta \to 0$, two additional, 
sharp peaks
  appear on both sides of the TZ and exponentially close to it
(peaks and TZ are visible only as a blip in Fig.\
  \ref{fig1}(i), due to lack of resolution) \cite{VF}. 
  Similar sharp features
  occur for $N>2$, see Fig.~\ref{fig2}(f).   
  These features, which get smeared out with increasing
  temperature, are irrelevant for the PL puzzle.
\bibitem{footnotetau} For $\tau > 0$, we calculate  
$|T|e^{i \alpha} \equiv - \int d \varepsilon \partial_\varepsilon 
f(\varepsilon) T^{\tau}_{\rm LR} (\varepsilon)$  
\cite{OregGefen97,twolevelpaper}, where the finite-temperature 
transmission matrix $T^{\tau}$ is obtained via the full density 
matrix NRG of \cite{Weichsel}.
\bibitem{U0remark} In a model with a weakly $V_g$-dependent, 
bare wide level and several narrow levels with $\delta >
  \Gamma$, Fano-like 
interference produces TZs and PLs even at $U=0$, for which 
interaction-induced occupation inversions are absent \cite{Yuval}.
\bibitem{twolevelpaper2} 
We have recently shown this explicitly for $N=2$ \cite{twolevelpaper}.
\end{thebibliography}
\end{document}